\documentstyle[12pt,aasms4]{article}
\begin{document}

\title{Harmonically dancing space-time nodes:\\
quantitatively deriving relativity, mass, and gravitation}

\author{Richard~Lieu$\,^{1}$}
\affil{\(^{\scriptstyle 1} \){Department of Physics, University of Alabama,
Huntsville, AL 35899, U.S.A.}
}

\begin{abstract}
The microscopic structure of space and time is investigated.
It is proposed that
space and time of an inertial observer $\Sigma$ are most conveniently
described as a crystal array $\Lambda$,
with nodes representing measurement `tickmarks' and
connected by independent quantized
harmonic oscillators which vibrate more severely the
faster $\Sigma$  moves with respect to
the object being measured (due to the Uncertainty
Principle).
The Lorentz transformation of Special Relativity is
derived.  Further, mass is understood as a localized region
$\Delta \Lambda$ having higher vibration 
temperature than that of the ambient
lattice.  The effect of relativistic mass increase may then be calculated
without appealing to energy-momentum conservation.  The origin of gravitation
is shown to be simply a
transport of energy from the boundary of $\Delta \Lambda$ outwards
by lattice phonon conduction, as the system tends towards
equilibrium.  Application to a single point mass leads readily to the
Schwarzschild metric, while a new solution is available for two point
masses - a situation where General Relativity is too complicated to
work with.  The important consequence is that inertial observers
who move at relative speeds too close to $c$ are no longer linked
by the Lorentz transformation, because the lattice of the `moving'
observer has already disintegrated into a liquid state.
\end{abstract}

\hspace{1cm} \\

In this work I propose a model for the microscopic structure of
space and time to understand relativity and the origin of
mass and gravitation in terms of statistical mechanics.

The underlying ideas originated from an examination of
Special Relativity (Einstein 1905),
which does not treat distances and times as
absolute.  For a uniformly `moving' observer,
one can disregard spatial dimensions
perpendicular to the velocity of motion ${\bf v}$ and elaborate the previous
statement as meaning that if $\Sigma'$ measures: (i) position differences
between events
at the same time, or (ii) time differences at the same position, the resulting
distances and intervals are smaller than
those of a `stationary' observer $\Sigma$ who measures
the same events from a frame at rest with respect to them, by a common factor
$\gamma = (1 - v^2/c^2)^{-\frac{1}{2}}$.
Since each of the two effects
(i) and (ii) was presented without interference
from the other, they can in general be
superposed to form the Lorentz transformation 
(i.e. rotation) of space-time (Lorentz 1923).
One could also describe the phenomenon as due to a measuring apparatus
having different intrinsic properties  when set in `motion'.   Thus
another way of expressing (i) is to visualize the
`tickmarks' which calibrate the ruler of $\Sigma'$
as more widely separated than those of $\Sigma$, as illustrated
in Figure 1.
If a unit of length for $\Sigma$ is $x_m$ then the same for
$\Sigma'$ will be $x_m \gamma$, which is of the
form $x_m \sqrt{N}$ ($N \geq 1$), suggesting
that microscopically space-time might be undergoing
a `random walk' process of some kind.  Similarly,
(ii) may also be viewed in this way, because the Lorentz
transformation of time is the same as that of space,
except time is being re-scaled by the speed
$c = 1/\sqrt{\epsilon_o \mu_o}$ to form
a different dimension, see Figure 2.  To avoid duplication, the ensuing
treatment will emphasize effect (i) only.

The particular manner of considering Special Relativity
as changes in the measuring devices
of a travelling observer matches well with a statistical approach:
an obvious way of further separating the `tickmarks' of a `moving' ruler,
is if they are not part of
a rigid body, but their relative positions can fluctuate while
in motion, maintaining the mean translational velocities $= {\bf v}$.
I shall pursue the fluctuations specifically as
oscillations, the amplitude of which
increase with the relative speed $v$.  The spirit of approach is
in accordance with the Heisenberg Uncertainty Principle
(Heisenberg 1930):
the higher the velocity (or momentum) being
measured the larger the inherent uncertainties
(deviations) in the ability of an apparatus to determine 
$p$ and $x$, with the
product $\Delta p \Delta x$ reaching the smallest value
$\sim \hbar$ only for ground states where motion of the object is minimized.
I therefore
let the natural separation of the `tickmarks'
be $x_o$ (note that $x_o < x_m$, for reasons to be explained).
To this
I add a sinusoidal
variable $x_1$, which has a mean $<x_1> = 0$.  The total
separation $x = x_o + x_1$, then, has a mean $<x> = x_o$, 
but the r.m.s. value is given by:
\begin{equation}
x_{rms} = x_o \left(1 + \frac{<x_1^2>}{x_o^2} \right)^{\frac{1}{2}}
\end{equation}
and is $> x_o$.  For the purpose of distance measurements it is
the magnitude of $x$, not the sign, which calibrates the length
between a pair of `tickmarks' at any phase of the cycle.
If $x$ oscillates rapidly\footnote{Caution should be
exercised while interpreting this word, because not only do
the spatial `tickmarks' oscillate, but the same applies to time
which behaves just like space
(see earlier).  Thus the period of a cycle cannot be defined with
respect to ordinary time.  The quantum mechanical approach of this
paper alleviates somewhat the urgency of settling the issue, however, since
one considers $<x^2>$ as an expectation value, proportional to energy.}
and substantially about
$x_o$, it is then $x_{rms}$ which defines a unit of length.  An
analogy is that of our mains AC power supply.
Most electric
appliances recognize only the magnitude of $V$ and have response times
slow compared with typical AC cycle
frequencies of $\sim$ a few $\times$ 10 Hz, so that the relevant
output voltage is
$V_{rms}$, even though $<V> = 0$.

The fundamental postulates of the theory are stated here.  The 
starting point concerns the first two of them.

\noindent
(a) Space {\it and} time form
a crystal lattice, the lattice points
(hereafter simply referred to as `nodes', which represent the
`tickmarks' of an observer) are connected by independent
harmonic oscillators with quantized energy levels
$E = (n + 1/2) \epsilon$.  
All inertial observers $\Sigma$ at
rest relative to each other are assigned
a common rest frame lattice $\Lambda$ with which measurements are 
naturally performed.

\noindent
(b) The degree of oscillation of $\Lambda$ is
parametrized by a temperature which may be defined not
only with respect
to $\Sigma$, but also {\it any} inertial observer $\Sigma'$.
If $\Sigma' = \Sigma$,
this will become the rest temperature, which
is taken to have an ambient value of 0 K.
If a relative speed $v \neq 0$ exists between the two frames, this
temperature will be $>$ 0 K, 
by an amount $T$ which increases with $v$ in a simple manner to
be formulated below; as a result
the quantity $x_{rms}$ is larger for $\Sigma'$ than for $\Sigma$.

\noindent
(c) Matter owes its origin
to a localized region $\Delta \Lambda$  of $\Lambda$
having a rest temperature $>$ 0 K.  The quantity we define
as rest {\it mass} is proportional to this 
surplus of energy within $\Delta \Lambda$.
An oscillator in $\Delta \Lambda$
has larger than ambient amplitude, i.e. $x_{rms} \rightarrow X_{rms}$.
With respect to $\Sigma'$ this amplitude is further increased, by
the same factor as $x_{rms}$ does in Postulate (b), see Figure 3.
The situation may be likened to $\Sigma$ being already in a local
Lorentz frame (which widens $x_{rms}$ to $X_{rms}$), 
and $\Sigma'$ moving relative to $\Sigma$ at the
stipulated speed $v$.

\noindent
(d) The effect of gravitation is due to conduction of energy by the
oscillator waves (phonons) from the hotter region of $\Delta \Lambda$
to the cooler ambient region.  The process causes a gradual downward
trend in the temperature as one moves away from $\Delta \Lambda$.
There is consequently a distribution of oscillator lengths, thereby
affecting the
separation of the measurement `tickmarks', see Figure 3.
This is the reason for the
curvature of space-time responsible for
the existence of universal gravitation in the lattice surrounding
$\Delta \Lambda$.

\noindent
None of the above is in conflict with the
First Relativity Postulate, viz. that no preferred
frame of rest exists.  There is also no violation of the Second
Postulate.

I now focus on Special Relativity, viz. Postulates (a) and (b).
When applying to inertial observers in relative motion, note that
even for $\Sigma$ the ambient separation between nodes of $\Lambda$
is larger than $x_o$, due to the zero point energy which leads
to a finite $<x_1^2>$ given by $<x_1^2> = E(n=0)/\kappa = \epsilon/2 \kappa$
where $\kappa$ is the oscillator (spring) constant.  The absolute
minimum unit of length is then obtained from Equ (1) as
\begin{equation}
x_m = x_o \left(1 + \frac{\epsilon}{2 \kappa x_o^2} \right)^{\frac{1}{2}}
\end{equation}
Since energy is additive, for finite $v$ (ambient $T > 0$) $x_m$
increases to:
\begin{equation}
x_{rms} = x_m \left(1 + \frac{<x_1^2>}{x_m^2} \right)^{\frac{1}{2}}
\end{equation}
where $<x_1^2> = \bar{E}(T) /\kappa$ is the mean energy at temperature $T$,
calculated with the lowest level now having energy $E = \epsilon$
(i.e. $n = 1$).  The relevant Partition Function is
$Z = 1/[1 - exp(-\epsilon/kT)]$, leading to a mean energy of
$\bar{E} = \epsilon Z e^{-\frac{\epsilon}{kT}}$.  Substituting
these into Equ (3), one obtains:
\begin{equation}
x_{rms} = x_m \left(1 + \frac{\epsilon}{2 \kappa x_o^2} 
\frac{e^{-\frac{\epsilon}{kT}}}{1-e^{-\frac{\epsilon}{kT}}}\right)^{\frac{1}{2}}
\end{equation}
At this point I complete postulate (b) above with the following
quantitative relationship between $v$ and $T$:
\begin{equation}
\frac{v^2}{c^2} = e^{-\frac{\epsilon}{kT}}~~ 
\left(= \frac{\bar{E}}{\epsilon + \bar{E}} \right)
\end{equation}
Combining (4) and (5), one gets:
\begin{equation}
x_{rms} = x_m \left(1 +
\frac{\epsilon}{2 \kappa x_o^2} \frac{\frac{v^2}{c^2}}{1 - \frac{v^2}{c^2}}
\right)^{\frac{1}{2}}
\end{equation}
Finally, postulate (a) is made more specific by invoking:
\begin{equation}
\epsilon = 2 \kappa x_o^2,
\end{equation}
which simply states that the zero point 
fluctuations double the mean-square separation
of nodes from the natural value of $x_o^2$.
Equs (4), (5) and (6) altogether tell us that the
unit length for observer $\Sigma'$ is not $x_m$, but $x_{rms} = x_m
\sqrt{1/(1-v^2/c^2)}$, implying smaller measured distances for
$\Sigma'$ than those for $\Sigma$, in accordance with
Lorentz contraction.

There remains the possibility that the suceess achieved so far
is an illusion, viz. statistical mechanics of harmonic oscillators
is a wrong and completely irrelevant approach to questions
concerning the nature of space and time.  Although it so happens that
Special Relativity can be explained in this way, the procedure could be
merely formal: a wide variety of mathematical formulae describe the
many diverse phenomena of the known physical world, the fact that some
of the formulae resemble each other in appearance does not immediately
imply a parallel in the physics involved.  In this instance, however, 
such concerns are settled by the theory's ability to go
beyond Special Relativity, to shed light on the issues of mass
and gravitation.

I first discuss the relativity of mass.  According to Postulate (c),
mass is an
energized  lattice region $\Delta \Lambda$ wherein an oscillator has 
large r.m.s. amplitude, as illustrated in Figure 2.
Upon transformation from $\Sigma$ to $\Sigma'$ (so that $\Lambda$ is
no longer the rest lattice) $X$, like $x_m$, is required by (c) and (b) to
increase by a factor $\gamma$.  Thus the oscillator energy is increased
by $\gamma^2$.  The number of oscillators is, of course, decreased by
$\gamma$, see Figure 2.  As a result, the total energy within $\Delta \Lambda$
is higher with respect to $\Sigma'$ by $\gamma$.  This then affords
an exceedingly simple derivation of the relativistic increase of the
energy within $\Delta \Lambda$.  By (c) the same effect applies to mass.

Next, I demonstrate how equally straightforward
it is to explain gravitation as an
energy transport effect.  Let us center $\Delta \Lambda$
at the origin of the rest lattice $\Lambda$, and consider the 1-D
conduction of energy in the +x-direction from some position
$x = x_{min}$, Figure 3.  Note that the notion of an 
inner  boundary from which
conduction commences is an abstract
one: the boundary does not have to represent the physical size of the
mass depicted in Figure .
The transport equation is:
\begin{equation}
\sigma_{th} \frac{dT}{dx} = n \bar{E} \bar{v}.
\end{equation}
Here $\sigma_{th} = n \bar{v} \lambda d\bar{E}/dT$ is the
thermal conductivity of phonons: $n$ is the linear phonon density,
$\bar{v}$ is the `speed of sound' in the lattice\footnote{The conduction
of energy takes place in $\Lambda$, which is a lattice of space {\it and}
time.  Thus, like the oscillations,
there is the need to introduce a new `time' axis when defining
the propagation speed of these phonons - 
signature of a fifth dimension.}, $\bar{E}(T)$ is the mean energy
of a phonon, and $\lambda$ is the phonon mean free path, which is
the size of the available lattice (since phonons do not interact),
i.e. $\lambda = x$.  Thus equ (8), together with the meaning of
the various symbols involved, imply that
\begin{equation}
-x \frac{d\bar{E}}{dx} = \bar{E},~~ \rm{or}~~ \bar{E} = \frac{1}{\alpha x}
\end{equation}
where $\alpha$ is a constant of integration.  Combining equs (5) and (9),
one obtains 
\begin{equation}
\frac{v^2}{c^2} = \frac{1}{\alpha \epsilon x +1}
\end{equation}
as equation giving the speed of the local Lorentz 
(or `free fall') frame at $x$.

Since the x-axis can represent a radial
direction, we now write $x = r - r_g$ and $r_g = x_{min}$.  In the limit
of $r \gg r_g$ ($v \ll c$) there should be agreement with Newtonian
gravity.  This requires $\alpha \epsilon = c^2/2GM$, which removes
the arbitrariness of the solution.  Returning to equ (9), one
clearly sees that $\bar{E} \propto M$, consistent with the
proposition in Postulate (c) that mass is proportional to
incumbent energy.  At high speeds the role of $r_g$ must be taken into
account ($v \rightarrow c$ as $r \rightarrow r_g$).
In the case of spherical symmetry there is only one
free parameter in the problem, i.e. $r_g$ must depend
on $\alpha$.  By setting $r_g = 1/\alpha \epsilon = 2GM/c^2$, equ (10)
reduces to $v^2/c^2 = r_g/r$, or $\gamma = (1 - \frac{r_g}{r})^{-\frac{1}{2}}$.
It is therefore apparent that the `tickmarks' of $r$ (or $x$) are
non-uniformly distributed throughout $\Lambda$, due to the energy 
outflow, in such a way that if they are used
to measure $r$ (or $x$) the result is
radius in Euclidean (flat) space-time.
In fact,  the expression for $\gamma$ contains all the information
one needs to construct the full Schwarzschild metric 
(Schwarzschild 1916) of General
Relativity.  For example, as a falling object approaches 
the gravitational radius (or event horizon) $r = r_g$
time dilation becomes infinite.  The notion of a `boundary' for
$\Delta \Lambda$ is now clear: phonon energy transport to other parts of
the lattice takes place only beyond the event horizon.

Apart from its ability to offer insightful, even elegant, explanations
of fundamental laws of physics, the model presented here can also make
predictions.  One immediate application is the problem of
two point masses, where no known solution of the Einstein Field
Equations exist.  The phonon conduction approach reduces the 
mathematical complexities by manifold.  A member of the pair
may be taken as the mass $M$ described above (except now
$M \rightarrow M_1$) while the other (mass $M_2$)
is positioned
at Euclidean distance $R$ from the first.  The separation between
their event horizons is then $R - r_g - r'_g$, where $r_g = 2GM_1/c^2$
and $r'_g = 2GM_2/c^2$.  Since the presence of mass $M_2$ cannot
alter the boundary conditions for the outward energy transport from
mass $M_1$\footnote{Otherwise one will not be able to restore the
single mass solution by letting the other mass tend to zero} and vice
versa, one simply solves for the oscillator (phonon) energy $E_1$
due to mass $M_1$, likewise $E_2$ due to $M_2$, for any position
along the line joining the two masses and distance $r$ from the
first.  The total energy at this position is 
$\bar{E} = \bar{E_1} + \bar{E_2} = \epsilon [ r_g/(r-r_g) + r'_g/(R-r-r'_g)]$.
By using equ (5), we have the speed $v$ of the local Lorentz frame at $r$:
\begin{equation}
\frac{v^2}{c^2} = \frac{\frac{r_g}{r-r_g} + \frac{r'_g}{R-r-r'_g}}
{1 + \frac{r_g}{r-r_g} + \frac{r'_g}{R-r-r'_g}}
\end{equation}
This gives for the first time
the space-time metric at any position between two masses.
For more complicated mass distributions, the metric at any point
is given by the superposition of all the outwardly conducted
energy distributions, in the same manner as above.

In conclusion, I found that a simple microscopic model of
space-time can explain the Lorentz transformation and the origin
of mass and gravitation very naturally.  In this model, space and
time form a crystal lattice, with nodes connected by harmonic
oscillators.  The notion of distances and times has
no meaning unless measurements are performed using the nodes
as `tickmarks' to define unit intervals.  The faster the speed
of an object, the `fuzzier' the measurement, and this is reflected
in equ (5) wherein the nodes vibrate with a temperature
$T$ which increases with the relative speed $v$ between a rest
lattice and a `moving' object.
When $v=c$,
$T$ reaches infinity, implying that the lattice has already
disintegrated before that.  Thus there must exist a threshold $v$ ($< c$)
which corresponds to a critical $T$, observers who move with respect
to each other at speeds high than this value are no longer connected
by the Lorentz transformation.  In other words, while Relativity
links the macroscopic properties of crystal lattices at
different vibration temperatures, here we are comparing two
fundamentally different lattices, viz. those of a crystal and a liquid.

I am indebted to Dr Massimilano Bonamente for suggesting a
harmonic oscillator model for the space-time lattice of a
moving observer.

\section*{Reference}

\noindent 
Einstein, A., 1905, {\it Annalen der Physik}, {\bf 18}, 891.

\noindent
Heisenberg, W., 1930, Physical Principles of the Quantum Theory,
Dover, NY.

\noindent
Lorentz, H.A., Einstein, A., Minkowski, H., \& Weyl, H., 1923, The Principles\\
\indent 
Relativity - a Collection of Original Memoirs, trans. W. Perret \& G.B. \\
\indent
Jeffrey, London: Methuen \& Co. Ltd., paperback reprint, Dover Publ. (1958).

\noindent
Schwarzschild, K., 1916, \"{U}ber das Gravitationsfeld eines Massenpunktes \\
\indent
nach der Einsteinschen Theorie, {\it Sitzber Preuss. Acad. Wiss. Berlin}\\
\indent
189 -- 196.

\newpage
 
\section*{Figure Captions}

\noindent 
Figure 1: The space-time lattice of stationary ($\Sigma$)
and moving ($\Sigma'$) observers are
illustrated here for the case of distance measurements.  The
`tickmarks' of the ruler of $\Sigma$ are marked as the topmost
set of black dots.  The rod to be measured is the short bar
immediately beneath, and is at rest with respect to $\Sigma$.
Observer $\Sigma'$ measures the length of this rod while in
motion, by simultaneously acquiring data on the positions of
the front and rear end of the rod.  It is postulated that
effectively $\Sigma'$ is using a moving set of `tickmarks',
and if microscopically these are connected by oscillators
which vibrate while in motion, the `tickmarks' widen as
depicted in the lower half of the diagram.  Consequently
$\Sigma'$ obtains a smaller value for the length of the rod.

\noindent
Figure 2: The space-time lattices of inertial observers $\Sigma$ (top)
and $\Sigma'$ (bottom), 
the latter `moving' at velocity ${\bf v}$ with respect
to the former, who is regarded as `stationary'.  If $\Sigma'$
measures $\Delta x$ between two `stationary' events at the
same time, or $\Delta t$ at the same position, in each case
the result is less than that obtained by $\Sigma$.  This means
for `orthogonal' measurements of space and time
performed by $\Sigma'$, the `tickmarks'
of $\Sigma$ are more widely separated, as indicated by the lower grid.

\noindent
Figure 3: Illustrating the origin of mass and gravitation.
{\it Top:} a region $\Delta \Lambda$ of $\Lambda$
(the rest frame lattice of observer $\Sigma$) has higher than
ambient temperature.  Mass is proportional to the
incumbent extra energy.
The physical boundary of the mass 
(shown here in the space dimension only) is
drawn as a rectangular box, inside of which all the
energy surplus resides, as a result the nodes are much more
widely spaced than outside.  {\it Middle:} The same region as it
appears in the lattice of observer $\Sigma'$ who `moves' with
respect to $\Sigma$ at velocity ${\bf v}$.  Separation between
any pair of nodes is now increased by the factor $\gamma$, meaning
fewer oscillators within $\Delta \Sigma$, but more energy per oscillator.
The net increase in enclosed energy (hence mass) is $\gamma$ (see text).
{\it Bottom:} Energy is conducted outwards from $\Delta \Lambda$ to
the ambient lattice by phonons, which causes the separation between
nodes to gradually reach the natural minimum at asymptotically large distances.
Note that the inner boundary from which the energy transport commences
is an abstract quantity which
need not be the physical size of the mass; in fact, the former is usually
within the latter.

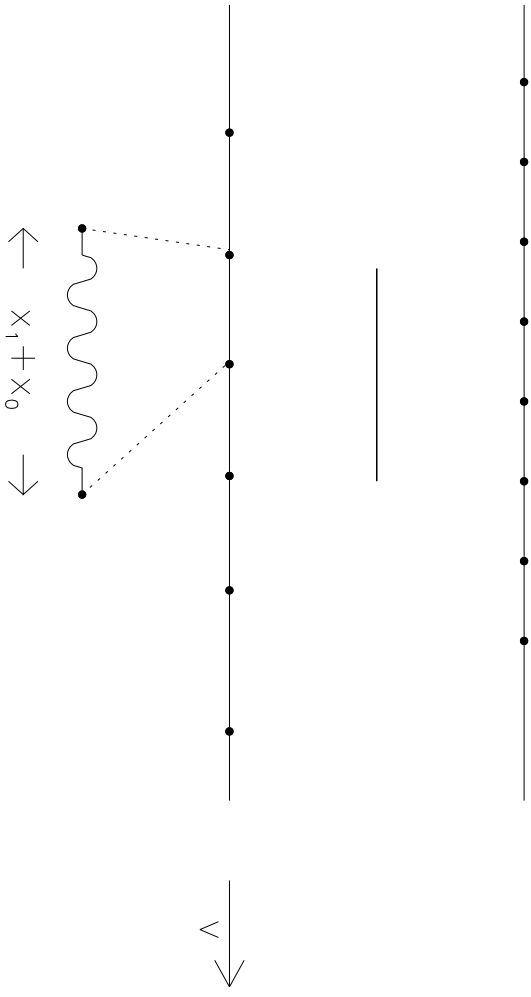
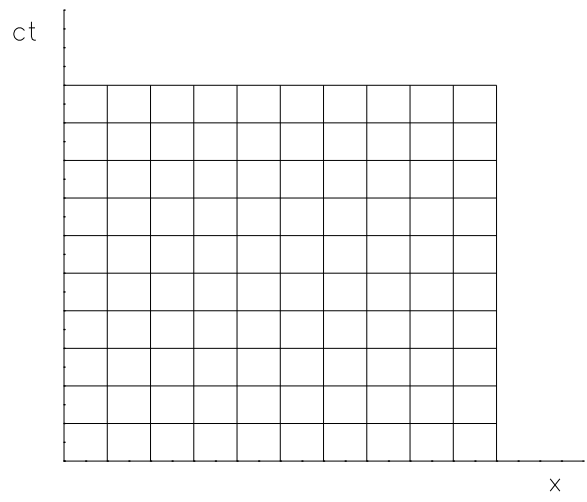
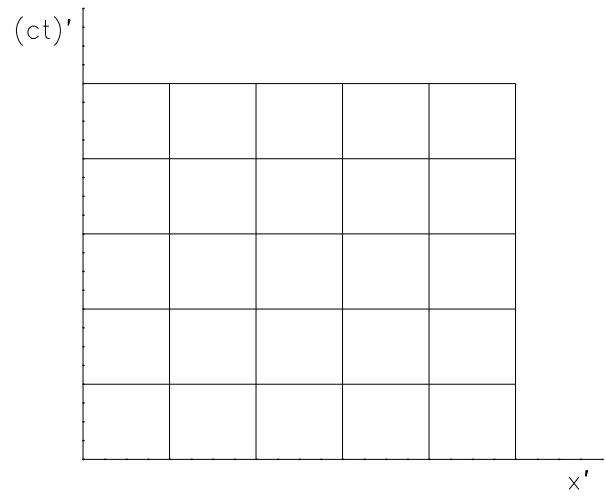
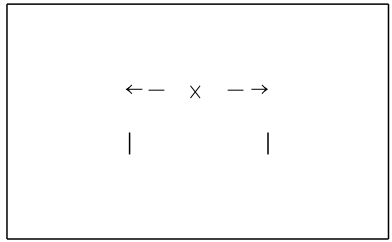
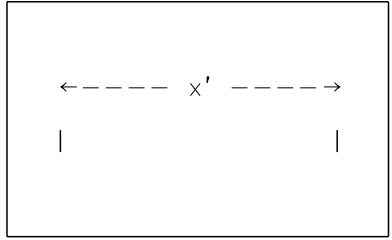
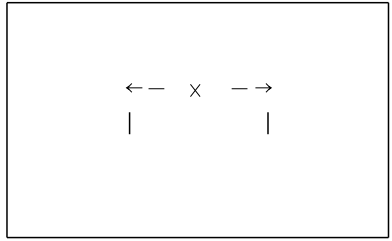
\end{document}